\documentclass[11pt,twocolumn]{article}

\usepackage[T1]{fontenc}
\usepackage[cp1250]{inputenc}
\usepackage{graphicx}
\usepackage{psfrag}
\usepackage{color} 
\usepackage{bbm}
\usepackage{amsmath}
\usepackage{hyperref}
\usepackage{authblk}
\usepackage[cm]{fullpage}
\usepackage[font=small,labelfont=bf]{caption}
\hypersetup{pdfstartview={FitH}}

\newcommand{\bq}{\begin{equation}}
\newcommand{\eq}{\end{equation}}
\newcommand{\ba}{\begin{eqnarray}}
\newcommand{\ea}{\end{eqnarray}}

\usepackage{xcolor}
\definecolor{TangoOrange3}{HTML}{CE5C00}

\begin{document}
 
\title{Mechanical interactions in bacterial colonies and the surfing probability of beneficial mutations}
\author[1]{Fred F. Farrell}
\author[2]{Matti Gralka}
\author[2,3]{Oskar Hallatschek}
\author[1,4]{Bartlomiej Waclaw}

\affil[1]{SUPA School of Physics and Astronomy, The University of Edinburgh, Mayfield Road, Edinburgh EH9 3JZ, UK}
\affil[2]{Department of Physics, University of California, Berkeley, CA 94720, USA}

\affil[3]{Department of Integrative Biology, University of California, Berkeley, California 94720, USA}
\affil[4]{Centre for Synthetic and Systems Biology, The University of Edinburgh, Edinburgh, UK}

\maketitle

\begin{abstract}
Bacterial conglomerates such as biofilms and microcolonies are ubiquitous in nature and play an important role in industry and medicine. 
In contrast to well-mixed, diluted cultures routinely used in microbial research, bacteria in a microcolony interact mechanically with one another and with the substrate to which they are attached. Despite their ubiquity, little is known about the role of such mechanical interactions on growth and biological evolution of microbial populations.
Here we use a computer model of a microbial colony of rod-shaped cells to investigate how physical interactions between cells determine their motion in the colony, this affects biological evolution. We show that the probability that a faster-growing mutant ``surfs'' at the colony's frontier and creates a macroscopic sector depends on physical properties of cells (shape, elasticity, friction). Although all these factors contribute to the surfing probability in seemingly different ways, they all ultimately exhibit their effects by altering the roughness of the expanding frontier of the colony and the orientation of cells. Our predictions are confirmed by experiments in which we measure the surfing probability for colonies of different front roughness. Our results show that physical interactions between bacterial cells play an important role in biological evolution of new traits, and suggest that these interaction may be relevant to processes such as {\it de novo} evolution of antibiotic resistance. 
\end{abstract}

\section{Introduction}
Bacteria are the most numerous organisms on Earth capable of autonomous reproduction. They have colonised virtually all ecological niches and are able to survive harsh conditions intolerable for other organisms such as high salinity, low pH, extreme temperatures, or the presence of toxic elements and compounds \cite{schlegel_general_1993}. Many bacteria are important animal or human pathogens, but some bacteria find applications in the industry as waste degraders \cite{pieper_engineering_2000} or to produce fuels and chemicals \cite{sabra_biosystems_2010}. In all these roles, biological evolution of microbes is an undesired side effect because it can disrupt industrial processes or lead to the emergence of new pathogenic \cite{chattopadhyay_high_2009} or antibiotic-resistant strains \cite{koch_evolution_2014}. 

Experimental research on bacterial evolution has been traditionally carried out in well-stirred cultures \cite{elena_evolution_2003,perron_rate_2008}. However, in their natural environment bacteria often form aggregates such as microcolonies and biofilms. Such aggregates can be found on food \cite{carpentier_biofilms_1993}, teeth (plague), on catheters or surgical implants \cite{costerton_bacterial_1999}, inside water distribution pipes \cite{berry_microbial_2006}, or in the lungs of people affected by cystic fibrosis \cite{gibson_pathophysiology_2003}. Bacteria in these aggregates  adhere to one another and the surface on which they live, form layers of reduced permeability to detergents and drugs, and sometimes switch to a different phenotype that is more resistant to treatment \cite{stewart_antibiotic_2001,drenkard_antimicrobial_2003,breidenstein_pseudomonas_2011}; this causes biofilms to be notoriously difficult to remove.

\begin{figure*}
\centering\includegraphics[width=0.85\textwidth]{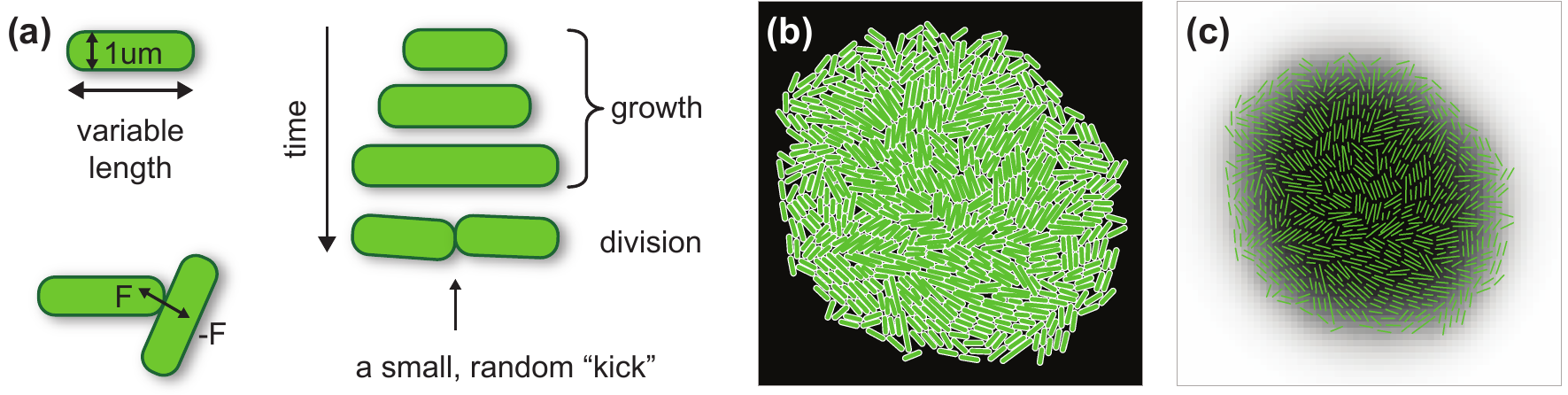}
\caption{\label{fig:algorithm}(a) Illustration of the computer algorithm. Bacteria are modelled as rods of varying length and constant diameter. When a growing rod exceeds a critical length, it splits into two smaller rods. (b) A small simulated colony. (c) The same colony with nutrient concentration shown as different shades of gray (white = maximal concentration, black = minimal); the cells are represented as thin green lines.}
\end{figure*}

An important aspect of bacteria living in dense conglomerates is that they do not only interact via chemical signaling such as quorum sensing \cite{singh_quorum_sensing_2000} but also through mechanical forces such as when they push away or drag other bacteria when sliding past them. Computer simulations \cite{boyer_buckling_2011,Farrell2013,giverso_emerging_2015,ghosh_mechanically_driven_2015} and experiments \cite{Volfson2008,su_bacterial_2012,asally_cover_2012,grant_role_2014,oldewurtel_differential_2015} have indicated that such mechanical interactions play an important role in determining how microbial colonies grow and what shape they assume. However, the impact of these interactions on biological evolution has not been explored. 

A particularly interesting scenario relevant to microbial evolution in microcolonies and biofilms is that of a range expansion \cite{klopfstein_fate_2006} in which a population of microbes invades a new territory. If a new genetic variant arises near the invasion front, it either ``surfs'' on the front and spreads into the new territory, or (if unlucky) it lags behind the front and forms only a small ``bubble'' in the bulk of the population~\cite{fusco2016excess}. This stochastic process, called ``gene surfing'', has been extensively studied \cite{Hallatschek2007,excoffier_genetic_2009, Hallatschek2010, behrman_species_2011, ali_reproduction_time_2012, Korolev2012, Lehe2012, gralka_allele_2016} but these works have not addressed the role of mechanical interactions between cells. Many of the existing models do not consider individual cells~\cite{Hallatschek2007}, assume Eden-like growth \cite{ali_reproduction_time_2012}, or are only appropriate for diluted populations of motile cells described by reaction-diffusion equations similar to the Fisher-Kolmogorov equation \cite{murray}. On the other hand, agent-based models of biofilm growth, which have been applied to study biological evolution in growing biofilms~\cite{kreft_biofilms_2004,Xavier2007,Xavier2009}, use very simple rules to mimic cell-cell repulsion which neglect important physical aspects of cell-cell and cell-substrate interactions such as adhesion and friction. 

In this work, we use a computer model of a growing microbial colony to study how gene surfing is affected by the mechanical properties of cells and their environment. In our model, non-motile bacteria grow attached to a two-dimensional permeable surface which delivers nutrients to the colony. This corresponds to a common experimental scenario in which bacteria grow on the surface of agarose gel infused with nutrients. We have previously demonstrated \cite{Farrell2013} that this model predicts a non-equilibrium phase transition between a regular (circular) and irregular (branched) shape of a radially expanding colony of microbes, and that it can be used to study biological evolution in microbial colonies \cite{gralka_allele_2016}. Here, we use this model to show that the surfing probability of a beneficial mutation depends primarily on the roughness of the expanding front of the colony, and to a lesser extend on the thickness of the front and cellular ordering at the front. We also investigate how mechanical properties of cells such as elasticity, friction, and adhesion affect these three quantities. We corroborate some of our results by experiments in which we vary the roughness of the growing front and show that it influences the surfing probability as expected.

\section{Computer model}
We use a computer model similar to that from Refs.\cite{Farrell2013,grant_role_2014,gralka_allele_2016}, with some modifications. Here we discuss only the generic algorithm; more details will be given in subsequent sections where we shall talk about the role of each of the mechanical factors.

We assume that bacteria form a monolayer as if the colony was two-dimensional and bacteria always remained attached to the substrate. This is a good approximation to what occurs at the edge of the colony and, as we shall see, is entirely justifiable because the edge is the part of the colony most relevant for biological evolution of new traits. We model cells as spherocylinders of variable length and constant diameter $d=2r_0=1\mu$m (Fig. \ref{fig:algorithm}a). Cells repel each other with normal force determined by the Hertzian contact theory: $F=(4/3)E r_0^{1/2} h^{3/2}$ where $h$ is the overlap distance between the walls of the interacting cells, and $E$ plays the role of the elastic modulus of the cell. The dynamics is overdamped, i.e. the linear/angular velocity is proportional to the total force/total torque acting on the cell:
\ba
	\frac{d\vec{r}_i}{dt} = \vec{F}/(\zeta m), \label{eq:drdt} \\
	\frac{d\phi_i}{dt} = \tau/(\zeta J).
\ea
In the above equations $\vec{r}_i$ is the position of the centre of mass of cell $i$, $\phi_i$ is the angle it makes with the $x$ axis, $\vec{F}$ and $\tau$ are the total force and torque acting on the cell, $m$ and $J$ are its mass and the momentum of inertia (perpendicular to the plane of growth), and $\zeta$ is the damping (friction) coefficient. We initially assume that friction is isotropic, and explore anisotropic friction later in Sec. \ref{sec:aniso}.

Bacteria grow by consuming nutrients that diffuse in the substrate. The limiting nutrient concentration dynamics is modelled by the diffusion equation with sinks corresponding to the bacteria consuming the nutrient:
\bq
	\frac{\partial c}{\partial t} = D \left( \frac{\partial^2 c}{\partial x^2} + \frac{\partial^2 c}{\partial y^2} \right) - k \sum_i \delta\left(\vec{r}_i - \vec{r}\right).
\eq
Here $\vec{r}=(x,y)$, $c=c(\vec{r},t)$ is the nutrient concentration at position $\vec{r}$ and time $t$, $D$ is the diffusion coefficient of the nutrient, and $k$ is the nutrient uptake rate. The initial concentration $c(\vec{r},0)=c_0$. 

A cell elongates at a constant rate $v_l$ as long as the local nutrient concentration is larger than a certain fraction (>1\%) of the initial concentration. When a growing cell reaches a pre-determined length, it divides into two daughter cells whose lengths are half the length of the mother cell. The critical inter-cap distance $l_{\rm cap-cap}$ at which this occurs is a random variable from a Gaussian distribution with mean $\ell_c$ and standard deviation $\pm 0.15\ \ell_c$. Varying $\ell_c$ allows us to extrapolate between quasi-spherical cells (e.g. yeasts {\it S. cerevisae} or the bacterium {\it S. aureus}) and rod-shaped cells (e.g. {\em E. coli} or {\em P. aeruginosa}), whereas the randomness of $l_{\rm cap-cap}$ accounts for the loss of synchrony in replication that occurs after a few generations (the coefficient of variation of the time to division $\sim 0.1-0.2$ \cite{hoffman_synchrony_1965,kennard_individuality_2016,iyer_biswas_scaling_2014}). The two daughter cells have the same orientation as the parent cell, plus a small random perturbation to prevent the cells from growing in a straight line. 

\begin{figure*}
\centering\includegraphics[width=0.85\textwidth]{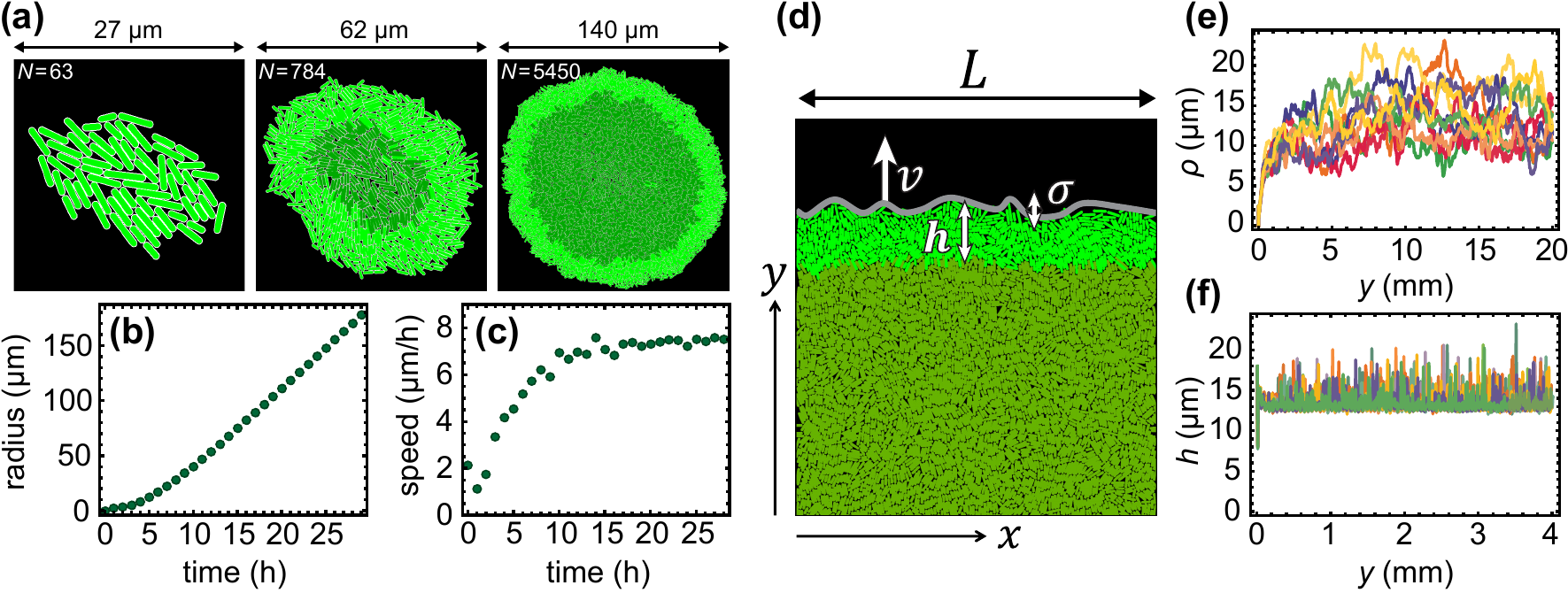}
\caption{{\bf (a)} Snapshots of a radially-growing simulated colony taken at different times (sizes), for $k=2$. Growing bacteria are bright green, quiescent (non-growing) bacteria are dark green. {\bf (b)} The radius of the colony increases approximately linearly in time. {\bf (c)} The expansion speed tends to a constant value for long times. {\bf (d)} Example configuration of cells from a simulation in a tube of width $L=80\mu$m. The colony expands vertically. $h$ is the thickness of the growing layer (Eq.~(\ref{eq:h})), $\rho$ is the roughness of the front (Eq.~(\ref{eq:sigma})). {\bf (e,f)} Thickness and roughness as functions of the position $y$ of the front, for $L=1280\mu$m and $k=2.5$, and for 10 indepedent simulation runs (different colours).}
\label{fig:round_growth}
\end{figure*}

We use two geometries in our simulations: a radially expanding colony that starts from a single bacterium (Fig. \ref{fig:round_growth}a), and a colony growing in a narrow (width $L$) but infinitely long vertical tube with periodic boundary conditions in the direction lateral to the expanding front (Fig. \ref{fig:round_growth}d). While the radial expansion case represents a typical experimental scenario, only relatively small colonies ($10^6$ cells as opposed to $>10^8$ cells in a real colony \cite{gralka_allele_2016}) can be simulated in this way due to the high computational cost.
The second method (growth in a tube) enables us to simulate growth for longer periods of time at the expense of confining the colony to a narrow strip and removing the curvature of the growing front. This has however little effect on the surfing probability of faster-growing mutants if the width $L$ of the tube is sufficiently large.

\begin{table}
	\centering
		\begin{tabular}{|c|c|c|}
			\hline
			Name & Value & Units \\
			\hline
			Nutrient diffusion constant $D$ & 50 & $\mu$m$^2$/h \\
			Nutrient concentration $c_0$ & 1 & a.u. \\
			Nutrient uptake rate $k$ & 1 -- 3 & a.u./h \\
			Young modulus $E$ & 100 & kPa \\
			Elongation length $v_l$ & 4 & $\mu$m/h \\
			Cell diameter & 1 & $\mu$m \\
			Average max. inter-cap distance $l_c$ & 4 & $\mu$m \\
			Damping coefficient $\zeta$ & 500 & Pa$\cdot$h \\
			\hline
		\end{tabular}
		\caption{\label{tab:params}Default values of the parameters of the model. This gives $\approx 30$min doubling time and the average length of bacterium $\approx 3\mu$m. If not indicated otherwise, all results presented have been obtained using these parameters.}
\end{table}

Figure \ref{fig:algorithm}b, shows a snapshot of a small colony; the concentration of the limiting nutrient is also shown. Table \ref{tab:params} shows default values of all parameters used in the simulation. Many of these parameters have been taken from literature data for the bacterium {\it E. coli} \cite{gralka_allele_2016}, but some parameters such as the damping coefficient must be estimated indirectly \cite{Farrell2013}. We note that the assumed value of the diffusion constant $D$ is unrealistically small; the actual value for small nutrient molecules such as sugars and aminoacids would be $\sim 10^6\mu$m$^2$/h, i.e., four orders of magnitude larger. Our choice of $D$ is a compromise between realism and computational cost; we have also showed in Ref.~\cite{Farrell2013} that the precise value of the diffusion coefficient is irrelevant in the parameter regime we are interested here. We also note that in reality cessation of growth in the center of the colony and the emergence of the growing layer may be due to the accumulation of waste chemicals, pH change etc., rather than nutrient exhaustion. Here we focus on the mechanical aspects of growing colonies and do not aim at reproducing the exact biochemistry of microbial cells, as long as the simulation leads to the formation of a well-defined growth layer 
(as observed experimentally).

\section{Experiments}
Experiments were performed as described in our previous work \cite{gralka_allele_2016}. Here we provide a brief description of these methods.

{\bf Strains and growth conditions.} For the mixture experiments measuring surfing probability, we used pairs of microbial strains that differed in fluorescence color and a selectable marker. The selective difference between the strains was adjusted as in \cite{gralka_allele_2016} using low doses of antibiotics. The background strains and antibiotics used were \textit{E. coli} DH5$\alpha$ with tetracycline, \textit{E. coli} MG1655 with chloramphenicol, and \textit{S. cerevisiae} W303 with cycloheximide. Selective differences were measured using the colliding colony assay \cite{Korolev2012}. 
\textit{E. coli} strains were grown on LB agar (2\%) medium (10g/L tryptone, 5g/L yeast extract, 10g/L NaCl) at either 37$^\circ$C or 21$^\circ$C. \textit{S. cerevisiae} experiments were performed on either YPD (20g/L peptone, 10g/L yeast extract, 20g/L glucose) or CSM (0.79g/L CSM (Sunrise media Inc.), 20 g/L glucose) at 30$^\circ$C. 20g/L agar was added to media before autoclaving. Antibiotics were added after autoclaving and cooling of the media to below 60$^\circ$C.

{\bf Measuring surfing probability.} For each pair of mutant and wild type, a mixed starting population was prepared that contained a low initial
 frequency $P_i$ of mutants having a selective advantage $s$. Colony growth was initiated by placing 2$\mu$l of the mixtures onto plates and incubated until  the desired final population size was reached. The initial droplet radius was measured to compute the number of cells at the droplet perimeter. The  resulting colonies were imaged with a Zeiss AxioZoom v16. The number of sectors was determined by eye. The surfing probability was calculated using Eq. (\ref{eq:psurfN}).

{\bf Timelapse movies.} For single cell-scale timelapse movies, we used a Zeiss LSM700 confocal microscope with a stage-top incubator to image the first few layers of most advanced cells in growing \textit{S. cerevisiae} and \textit{E. coli} colonies between a coverslip and an agar pad for about four hours, taking an image every minute.

{\bf Measuring roughness.} Images of at least 10 equal-sized colonies per condition were segmented and the boundary detected. The squared radial distance $\delta r^2$ between boundary curve and the best-fit circle to the colony was measured as a function of the angle and averaged over all possible windows of length $l$. The resulting mean $\delta r^2$ was averaged over different colonies. 

Images of moving fronts at the single-cell level from the timelapse movies were first segmented using a local adaptative threshold algorithm to identify cells. The front was found by the outlines of cells directly at the front. For all possible windows of length $l$, a line was fitted to the front line and the mean squared distance from the best-fit line was measured, as in Ref.~\cite{Hallatschek2007}. The resulting mean squared distance was averaged over all windows of length $l$ and all frames.

\section{Simulation results}

\subsection{Growth and statistical properties of the simulated colony\label{sec:statprop}}
We now discuss the properties of our simulated colonies. When the colony is small, all bacteria grow and replicate. As the colony expands, the nutrient becomes depleted in the centre of the colony because diffusion of the nutrient cannot compensate its uptake by growing cells. This causes cessation of growth in the centre. When this happens, growth becomes restricted to a narrow layer at the edge of the colony, see Fig. \ref{fig:round_growth}a, and Supplementary Video 1. The radius of the colony increases approximately linearly in time (Fig. \ref{fig:round_growth}b,c). The presence of a ``growing layer'' of cells and the linear growth of the colony's radius agree with what has been observed experimentally \cite{freese_genetic_2014,gralka_allele_2016}.

Statistical properties of the growing layer can be conveniently studied using the ``tube-like'' geometry. Figure \ref{fig:round_growth}d shows a typical configuration of cells at the colony's frontier (see also Supplementary Video 2). The growing layer can be characterized by its thickness $h$ and roughness $\rho$ which we calculate as follows. We first rasterize the growing front of the colony using pixels of size $1\times 1\mu$m, and find the two edges of the front: the upper one (the colony edge) $\{y_i^+\}$ and the lower one (the boundary between the growing and quiescent cells) $\{y_i^-\}$. We then calculate the average thickness as
\bq
	h = \frac{1}{L} \sum_{i=1}^L  \min_{j=1,\dots,L} \sqrt{(i-j)^2+(y_i^+ - y_j^-)^2}. \label{eq:h}
\eq
This method takes into account that the growing layer can be curved and does not have to run parallel to the $x$ axis\footnote{Alternatively, $h$ can be defined as the area of the colony that contains replicating cells divided by the interface length $L$. Both methods produce similar results.}. Similarly, we calculate the average roughness as
\bq
	\rho = \sqrt{\frac{1}{L} \sum_{i=1}^L (y_i^+ - Y^+)^2} \;, \label{eq:sigma}
\eq
where $Y^+ = (1/L)\sum_i y_i^+$. Note that all quantities ($L,Y^+,y_i^+,y_i^-$) are in pixels and not $\mu$m.

\begin{figure}
\centering
 \includegraphics[width=\columnwidth]{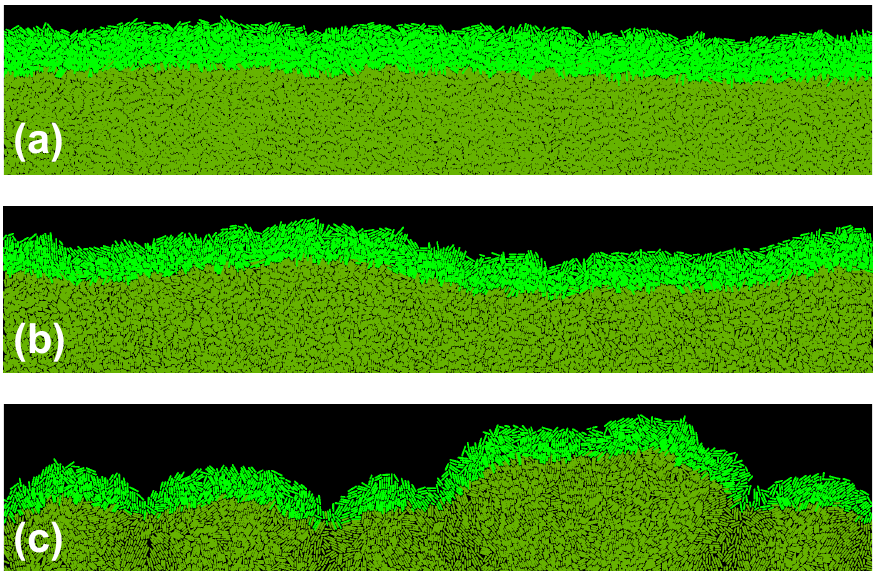}
\caption{\label{snapshots}The frontier of the colony for three different nutrient uptake rates $k=1.8$ (a), $k=2.2$ (b) and $k=2.6$ (c). The thickness of the growing layer (bright green) decreases only moderately ($1.64\times$) from $h=13.5\pm 0.1\mu$m for $k=1.8$ to $h=8.2\pm 0.1\mu$m for $k=2.6$, but this has a large impact on the front roughness which changes from $\rho=2.1\pm 0.2\mu$m to $\rho=9.3\pm 0.4\mu$m, correspondingly. For $k=2.6$ the growing layer begins to loose continuity and splits into separate branches.}
\end{figure}

After a short transient the expansion velocity, the nutrient profile, and other properties of the growing layer stabilize and vary little with time (Fig. \ref{fig:round_growth}e,f). It is therefore convenient to choose a new reference frame co-moving with the leading edge of the colony. Since cells that lag behind the front do not replicate, we do not have to simulate these cells explicitly. This dramatically speeds up simulations and enables us to study stripes of the colony of width $L>1$mm and length $>10$mm. 

\begin{figure*}
\centering\includegraphics[width=0.85\textwidth]{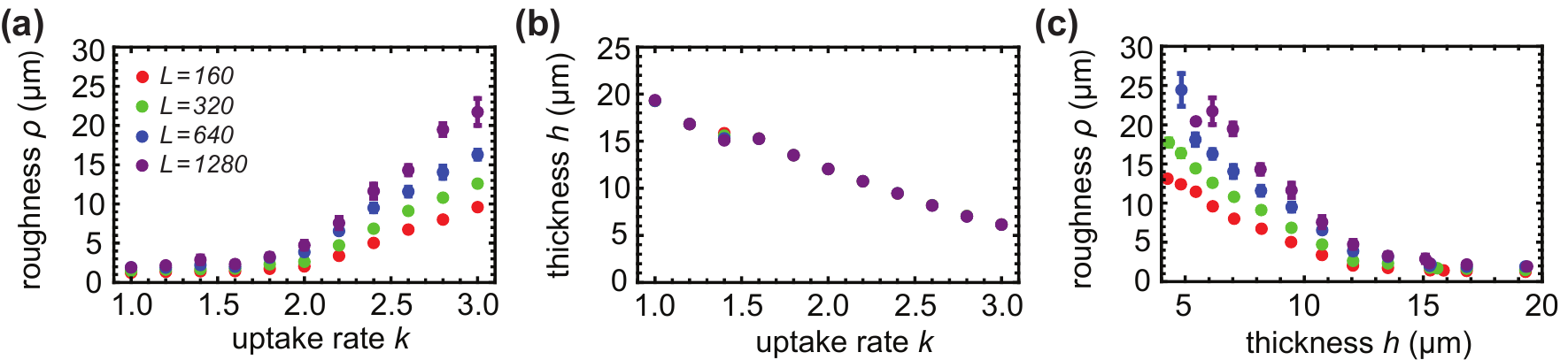}
\caption{\label{h_and_sigma_diff_k}Thickness and roughness of the growing layer for different front lengths (tube widths) $L=160$ (red), $L=320$ (green), $L=640$ (blue), and $L=1280\ \mu$m (purple). (a) Thickness $h$ decreases as the nutrient uptake rate $k$ increases. $h$ does not depend on the length $L$ of the front. (b) Roughness $\rho$ increases with both $k$ and $L$. (c) Roughness versus thickness; different points correspond to different $k$ from the left and middle figure. }
\end{figure*}

We have shown previously \cite{Farrell2013} that the thickness of the growing layer of cells is controlled by the nutrient concentration $c_0$, nutrient uptake rate $k$, growth rate $b$, and elasticity $E$ of cells. This in turn affects the roughness of the leading edge of the colony, see Fig.~\ref{snapshots}, where we vary the uptake rate $k$ while keeping the remaining parameters constant. 
Figure \ref{h_and_sigma_diff_k} shows that front thickness decreases and its roughness increases with increasing $k$; eventually, when a critical value $k_c\approx 2.5$ is crossed, the growing front splits into separate branches. This transition has been investigated in details in Ref. \cite{Farrell2013}. Although this scenario can be realized experimentally \cite{Fujikawa1989,Kawasaki1997}, here we focus on the ``smooth'' regime in which colonies do not branch out and the frontier remains continuous.

\subsection{Surfing probability of a beneficial mutation}
\begin{figure}
\includegraphics{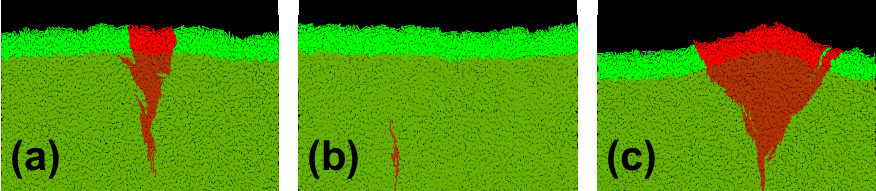}
\caption{The fate of mutants. Left and middle panels show different fates of a sector of fitter ($s=0.1$) mutant cells (red) in a colony of ``wild-type'' cells (green). The sector can either expand (left panel) or collapse and become trapped in the bulk when random fluctuations cause mutant cells to lag behind the front (middle panel). Right panel shows a sector with larger ($s=0.5$) growth advantage; significantly faster growth of mutant cells leads to a ``bump'' at the front. In all cases $k=1.8, L=160\mu$m.}\label{sector_shots}
\end{figure}

When a mutation arises at the colony's frontier, its fate can be twofold \cite{Hallatschek2007,gralka_allele_2016}. If cells carrying the new mutation remain in the active layer, the mutation ``surfs'' on the moving edge of the colony and the progeny of the mutant cell eventually forms a macroscopic ``sector'' (Fig. \ref{sector_shots}). On the other hand, if cells carrying the mutation leave the active layer, the mutation becomes trapped as a bubble in the bulk of the colony~\cite{fusco2016excess}. Due to the random nature of replication and mixing at the front, surfing is a stochastic process; a mutation remains in the active layer in the limit $t\to\infty$ with some probability $P_{\rm surf}$ which we shall call here the surfing probability. 

Surfing is a softer version of fixation - a notion from population genetics in which a mutant takes over the population. The soft-sweep surfing probability has therefore a hard-selection-sweep counterpart, the fixation probability, which is the probability that the new mutation spreads in the population so that eventually all cells have it. Both surfing and fixation probabilities depend on the balance between selection (how well the mutant grows compared to the parent strain) and genetic drift (fluctuations in the number of organisms due to randomness in reproduction events)~\cite{M.A.Nowak2006}. In the previous work \cite{gralka_allele_2016} we showed that $P_{\rm surf}$ increased approximately linearly with selective advantage $s$ -- the difference between the growth rate of the mutant and the parent strain. Here, we study how the properties of the active layer affect $P_\mathrm{surf}$ for a fixed $s$. 

We first run simulations in the planar-front geometry in which a random cell picked up from the growing layer of cells with probability proportional to its growth rate is replaced by a mutant cell with selective advantage $s>0$. This can be thought of as mutations occurring with some small probability per division. The simulation finishes when either fixation (all cells in the growing layers becoming mutants) or extinction (no mutant cells in the growing layer) is achieved. Before inserting the mutant cell, the colony is simulated until the properties of the growing layer stabilize and both thickness and roughness reach steady-state values. The simulation is then repeated many times and the probability of surfing is estimated from the proportion of runs leading to fixation of the mutant in the growing layer. Snapshots showing different fates (extinction, surfing) of mutant sectors are shown in Fig~\ref{sector_shots}.

\begin{figure}
\includegraphics{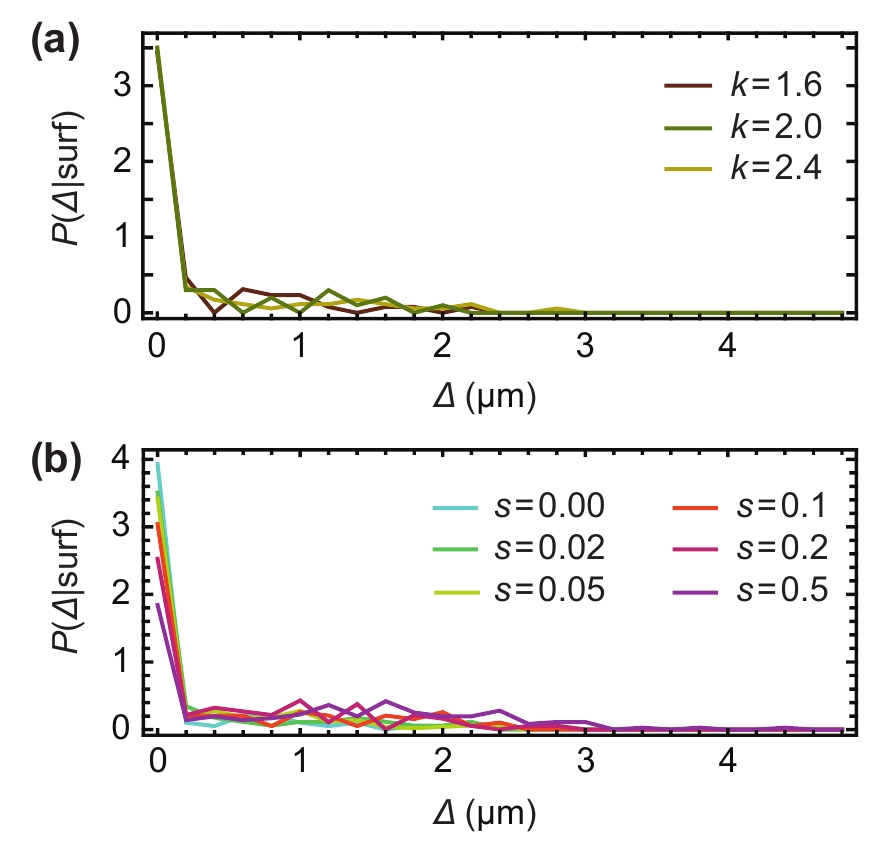}
\caption{\label{pzgiven} (a) $P(\Delta|\textrm{surf})$ for $L=160\mu$m, selective advantage $s=0.02$, and different $k=1.6,2.0,2.4$.
(b) $P(\Delta|\textrm{surf})$ for $L=160\mu$m, $k=2.0$, and different selective advantages $s=0,0.02,0.05,0.1,0.2,0.5$. Only mutants from the first layer of cells have a significant chance of surfing.}
\end{figure}

{\bf Surfing probability depends on the position of the cell in the growing layer.}
In Ref. \cite{gralka_allele_2016} we showed that the surfing probability strongly depends on how deeply in the growing layer a mutant was born. Here we would like to emphasize this result as it will become important later. Let $\Delta$ be the distance from the edge of the colony to the place the mutant first occurred. Figure~\ref{pzgiven} shows the probability density $P(\Delta|\textrm{surf})$ that a cell was born
a distance $\Delta$ behind the colony front, \emph{given that it went on to surf on the edge of the expanding colony}. It is evident that only cells born extremely close to the frontier have a chance to surf. Cells born deeper must get past the cells in front of them. This is unlikely to happen, even if the cell has a significant growth advantage, as the cell's growth will also tend to push forward the cells in front of it. This also justifies why we focus on 2d colonies; even though real colonies are three-dimensional, all interesting dynamics occurs at the edge of the colony, made of a single layer of cells. 

Given that surfing is restricted to the first layer of cells, and the distribution $P(\Delta|\textrm{surf})$ is approximately the same for all explored parameter sets (different $k$ and $s$), it may seem to be a waste of computer time to study the fate of mutants that occurred deeply in the growing layer. To save the time, and to remove the effect the front thickness has on $P_{\rm surf}$ (thicker layer = lower overall probability), we changed the way of introducing mutants. Instead of inserting mutants anywhere in the growing layer, we henceforth inserted them only at the frontier.

\begin{figure}
\centering\includegraphics[width=\columnwidth]{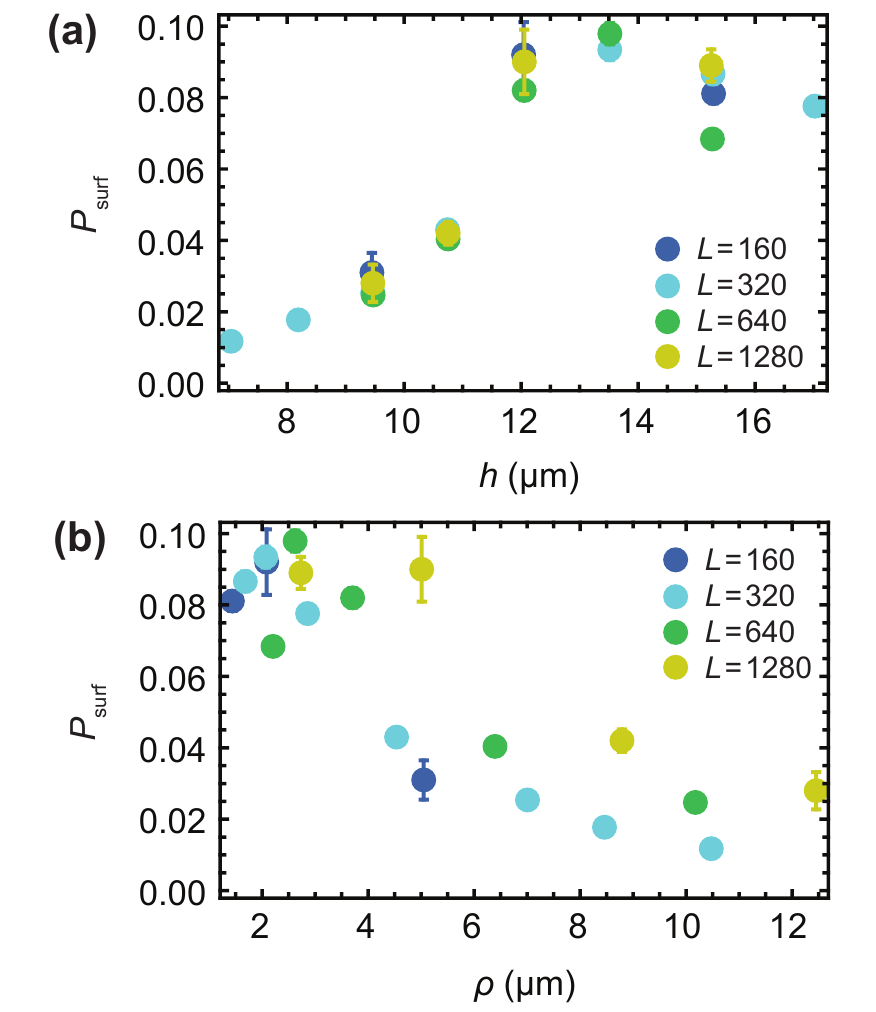}
\caption{\label{psurf} (a) $P_{\rm surf}$ for different thickness $h$ of the growing layer, for $s=0.02$ and $L=160, 320, 640, 1280\ \mu$m (different colours). (b) the same data as a function of front roughness $\rho$. Between $10^3$ and $10^4$ simulations were performed for each data point to estimate $P_{\rm surf}$.}
\end{figure}

{\bf Roughness of the front is more predictive of $P_{\rm surf}$ than its thickness.} 
Using the new method of introducing mutants (only the first layer of cells), we run simulations for $s=0.02$ and for different widths $L$ and nutrient uptake rates $k$ as in Fig. \ref{h_and_sigma_diff_k}. Figure \ref{psurf} shows how the surfing probability $P_{\rm surf}$ varies as a function of the thickness and the roughness of the front. $P_{\rm surf}$ increases with increasing thickness $h$ and decreases with increasing roughness $\rho$. We know from Fig. \ref{h_and_sigma_diff_k} that thickness and roughness are inversely correlated so this reciprocal behaviour is not surprising. An interesting question is which of the two quantities, roughness or thickness, directly affects the probability of surfing? From a statistics point of view, thickness $h$ seems to be a better predictor of $P_{\rm surf}$ because data points for the same $h$ but for different $L$ correlate better. However, it could be that it is actually front roughness that directly (in the causal sense) affects the surfing probability and that $P_{\rm surf}$ and $h$ are anti-correlated because of the relationship between $h$ and $\rho$.

\begin{figure}
\centering\includegraphics[width=0.95\columnwidth]{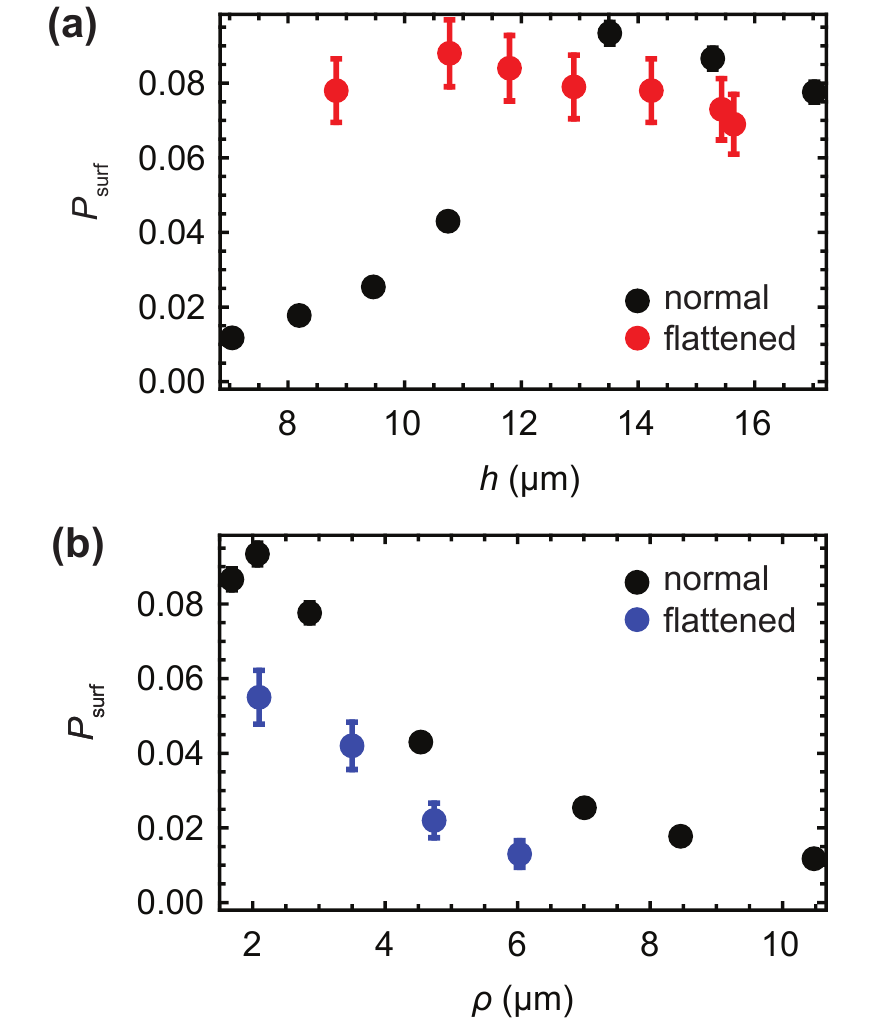}
\caption{\label{fig:psurf_flat_front} (a) $P_{\rm surf}$ as the function of front thickness $h$ for the normal (black) and flattened front (red, $g=500$), for $L=320\mu$m. We vary the nutrient uptake rate $k=1.6...2.8$ to simulate fronts of different thickness. The flat front has roughness $\rho$ between $0.84$ and $1.0$ for all $k$.
 (b) $P_{\rm surf}$ for the normal (black) and flattened front (blue) as the function of roughness $\rho$. The flattened front has approximaly the same thickness for all data points ($h$ between $10.0$ and $10.3\mu$m). The points correspond to maximum roughness set to $\rho_{\rm max}=2, 3.5, 5$, and $7$, for $k=2.6$; the actual (measured) $\rho$ differs very little from these values.
}
\end{figure}

We performed two computer experiments to address the above question. First, we simulated a colony that had a very low and constant roughness $\rho\approx 1$, independently of front's thickness. This was achieved by introducing an external force $F_y=-gy$ acting on the centre of mass of each cell, where $g>0$ was a ``flattening factor'' whose magnitude determined the strength of suppression of deviations from a flat front. $P_{\rm surf}$ plotted in Figure \ref{fig:psurf_flat_front}, left, as a function of $h$ for two cases: ``normal'', rough front, and ``flattened'' front, demonstrates that the surfing probability does not depend on $h$ in the case of flat front. 

Second, we varied roughness while keeping thickness constant. This was done by measuring front roughness in each simulation step, and switching on
 the external ``flattening'' force $F_y=-gy$ if the roughness was larger than a desired value $\rho_{\rm max}$. Figure \ref{fig:psurf_flat_front}, right, shows that although thickness remains the same for all data points, $P_{\rm surf}$ decreases with increasing roughness.

We can conclude from this that it is the increase in the roughness, and not decreasing thickness, that lowers the surfing probability for thinner fronts (larger nutrient intake rate $k$). However, the data points in Fig. \ref{psurf}, right, from different simulations do not collapse onto a single curve as it would be expected if average, large-scale front roughness was the only factor.

\begin{figure*}
\centering\includegraphics[width=0.95\textwidth]{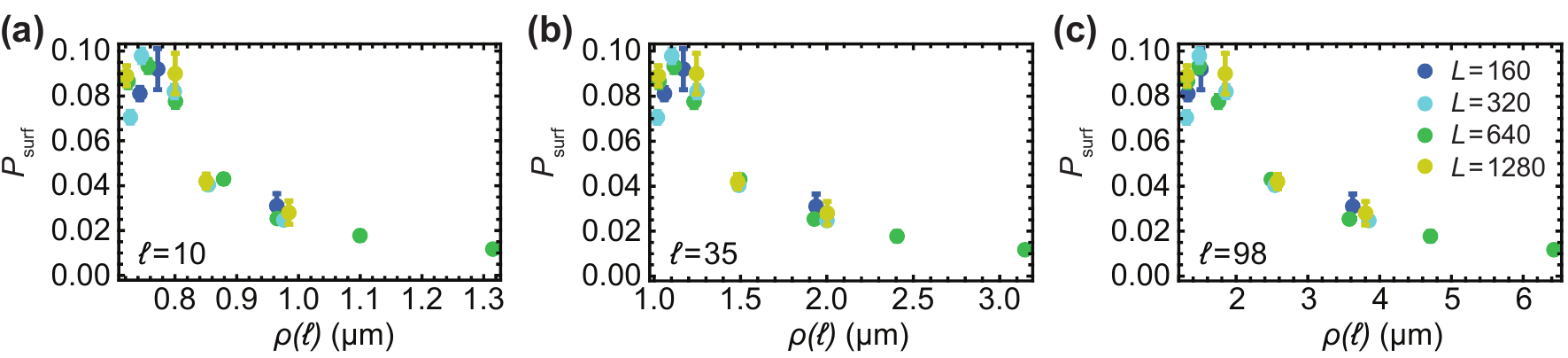}
\caption{\label{fig:pfix_rho}$P_{\rm surf}$ as the function of local roughness $\rho(l)$ of the growing layer, for different sizes $L=160,320,640,1280\ \mu$m (as in Fig. \ref{psurf}) and $s=0.02$. Left: $l=10$, middle: $l=35$, right: $l=98\ \mu$m. For each $l$, data points for different $L$ collapse onto a single curve.}
\end{figure*}

{\bf Local roughness predicts $P_{\rm surf}$}.
According to the theory of Ref. \cite{Hallatschek2010}, the dynamics of a mutant sector can be described by a random process similar to Brownian motion in which the sector boundaries drift away from each other with constant velocity. The velocity depends on the growth advantage $s$ whereas the amplitude of random fluctuations in the positions of boundary walls is set by the microscopic dynamics at the front. We reasoned that these  fluctuations must depend on the roughness $\rho$ of the frontier, and that a mutant sector should be affected by front roughness when the sector is small compared to the magnitude of fluctuations. This means that local roughness $\rho(l)$, determined over the length $l$ of the front, should be more important than the global roughness $\rho(L)$.
We calculated the local roughness as
\bq
	\rho(l) = \frac{1}{n} \sum_{i=1}^{n} \sqrt{\frac{1}{l} \sum_{j=i}^{i+l} (y_j^+ - Y^+)^2}. \label{eq:locsigma}
\eq
Here $Y^+$ is the average height of the interface and $\{y_i^+\}$ are the vertical coordinates (interface height) of the points at the leading edge, obtained as in Section \ref{sec:statprop}. Figure \ref{fig:pfix_rho} shows that $P_{\rm surf}$ for different $L$ now collapse onto a single curve, for all lengths $l\approx 10\dots 100\mu$m over which roughness has been calculated.

\begin{figure*}
\centering\includegraphics[width=0.95\textwidth]{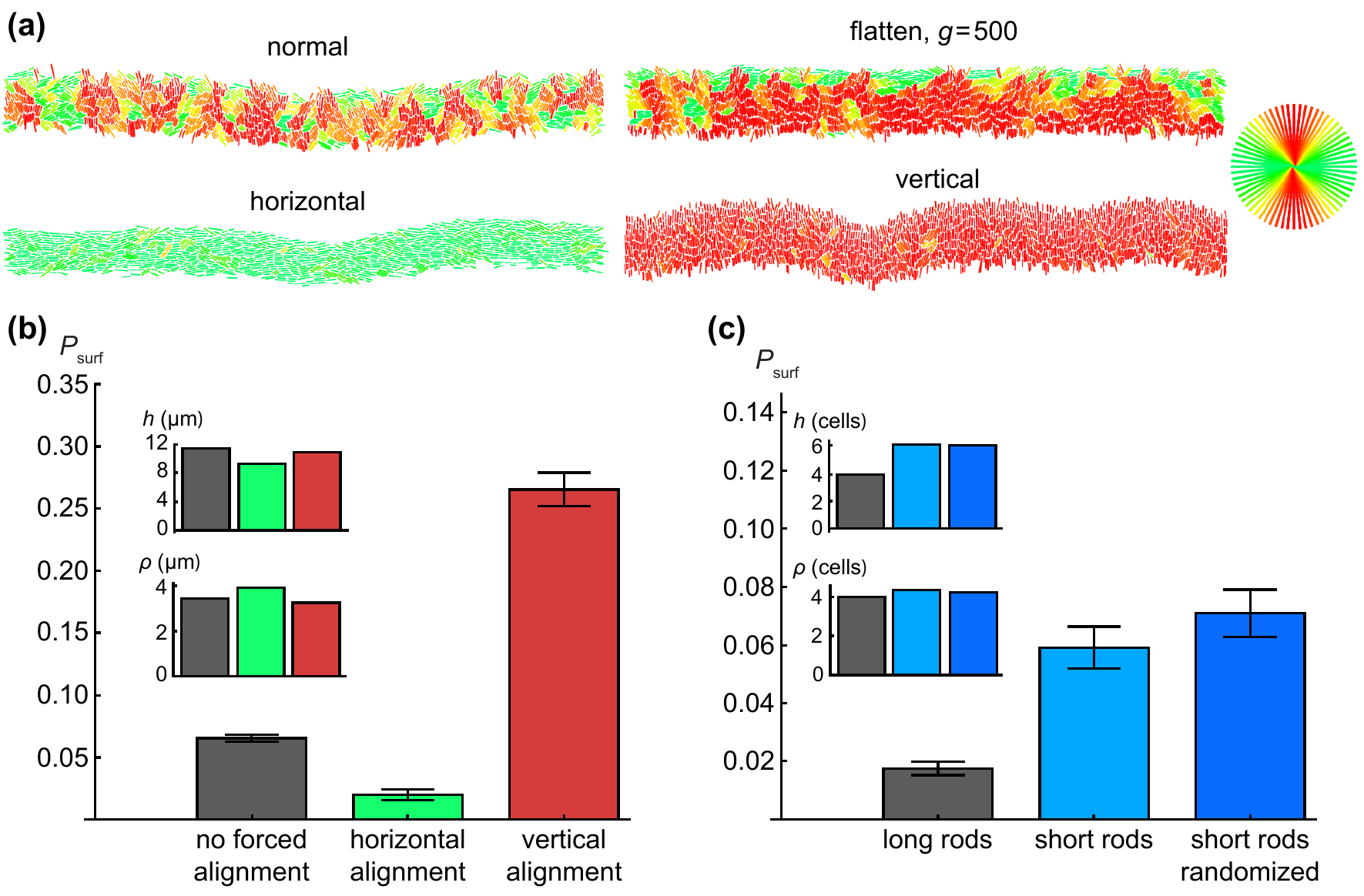}
\caption{\label{fig:domains_and_psurf} (a) Orientation of cells (colours as in the circle in the upper-right corner) in the growing layer for different models. (b, c) Comparison of fixation probabilities for different cellular alignments at the front, for approximately the same thickness and roughness, both of which were controlled by varying $k$. To achieve this, different $k$ needed to be used in panels (b, c) and hence the two panels cannot be directly compared. In all cases $L=320\mu$m, $s=0.02$. For horizontally- and vertically-forced cells, $\tau_{\rm max}=10000$. Short cells have a maximum length of $2\mu$m; upon division, they become circles of diameter $1\mu$m.}
\end{figure*}

{\bf Orientation of cells affects $P_{\rm surf}$.}
So far we have focused only on the macroscopic properties of the leading edge of the colony, completely neglecting its granular nature due to the presence of individual cells. Recall that in our model each cell is rod-shaped, and the direction in which it grows is determined by the orientation of the rod. Figure \ref{fig:domains_and_psurf}a shows that cells at the leading edge assume orientations slightly more parallel to the direction of growth (vertical) in the flattened front than in the normal simulation. A natural question is how does cellular alignment affects $P_{\rm surf}$, independently of the roughness? To answer this question, we simulated a modified model, in which external torque $\tau=-\tau_{\rm max} \sin[(\phi - \phi_{\rm preferred})\mod \pi]$ was applied to the cells, forcing them to align preferentially in the direction $\phi_{\rm preferred}$. We investigated two forced alignments: $\phi_{\rm preferred}=0$ corresponding to cells parallel to the $x$ axis and hence to the growing edge of the colony, and $\phi_{\rm preferred}=\pi/2$ which corresponds to the vertical orientation of cells (perpendicular to the growing edge). 

Figure \ref{fig:domains_and_psurf}b compares these two different modes with previous simulations with no external torque, for approximately the same thickness and roughness of the growing layer. It is evident that the orientation of cells strongly affects the surfing probability: horizontally-forced cells have $\sim 3$x smaller $P_{\rm surf}$ compared to the normal case, which in turn has $P_{\rm surf}$ $\sim 5$x smaller than vertically-forced cells.

{\bf Shorter cells have higher $P_{\rm surf}$ than long cells.}
To check how the aspect ratio of cells affect $P_{\rm surf}$, we simulated cells whose maximal length was only $2\mu$m and the minimal separation before the spherical caps was zero, i.e., the cells became circles immediately after division. As before we selected a set of $k$'s such that the thickness and roughness were approximately the same for all simulations. In order to make a fair comparison between ``short rods'' and ``long rods'' from previous simulations, thickness and roughness were expressed in cell lengths rather than in $\mu$m. This was done by dividing both $h$ and $\rho$ by the average length of a cell measured for cells from the growing layer. Figure \ref{fig:domains_and_psurf}c show that short rods have a much higher surfing probability than long rods.

In all previous simulations, even for short rods, cells remembered their orientation from before division and growth always initially occurred in that direction. To see whether this has any impact on $P_{\rm surf}$, we considered a scenario in which the new direction of growth is selected randomly and does not correlate with the direction prior to division. Figure \ref{fig:domains_and_psurf}c shows that $P_{\rm surf}$ almost does not change regardless whether a short cell randomly changes its orientation after division or not.

\subsection{Surfing probability and the mechanical properties of bacteria\label{sec:aniso}}
Our results from the previous section demonstrate that surfing is affected by (i) the roughness of the growing layer, (ii) the orientation of cells, (iii) the thickness of the growing layer if mutations occur inside the growing layer and not only at its edge. To show this, we varied thickness, roughness, and orientation of cells by using {\it ad hoc} external forces flattening out the front or forcing the cells to order in a particular way. In this section we will investigate what parameters of the model affect surfing in the absence of such artificial force fields.

{\bf Thickness of the growing layer.} If cells are prohibited to form multiple layers, as in our 2d simulations, thickness $h$ can be determined from the parameters of the model by a simple dimensional analysis. Assuming that $h$ is proportional to the characteristic scale over which the nutrient concentration and cell density reaches bulk values~\cite{Farrell2013}, we can approximate $h$ by
\bq
	h \approx \sqrt{\frac{E}{(\zeta/a)\phi}}(1/\beta-1)^{3/4}, \label{eq:happrox}
\eq
where $E$ is the elastic modulus of the bacterium (Pa), $a$ is the average area per cell ($\mu$m$^2$), $\zeta$ is the friction coefficient (Pa$\cdot$h), $\phi$ is the replication rate (h$^{-1}$), and $\beta<1$ is a dimensionless ratio of the nutrient consumption rate to biomass production rate (i.e. new bacteria): $\beta=(k\rho_0)/(\phi c_0)$. Equation (\ref{eq:happrox}) shows that thickness $h$ increases with increasing cell stiffness (larger $E$) and replication rate $\phi$, and decreases with increasing nutrient uptake $k$ and increasing friction $\zeta$. The aspect ratio of the cells does not affect $h$ in our model. Equation (\ref{eq:happrox}) suggests that the thickness of the growing layer can be conveniently controlled in an experiment by varying temperature or growth medium (which both affect the growth rate), or by varying the nutrient concentration $c_0$. We shall use the first two methods when discussing the experimental verification of our theory.

{\bf Orientation of cells.}
A useful measure of the global alignment of cells in the colony is the order parameter $S=\left<\cos^2(\phi-\Phi)\right>$. Here $\phi$ is the angle a cell makes with the $x$-axis and $\Phi$ is the angular coordinate of the vector normal to the front; this is to remove a trivial contribution to $S$ due to the curvature of the front caused by roughness. According to this definition, $S=1$ if all cells are perfectly vertically aligned (in the direction of growth), $S=0$ if they are horizontal (parallel to the front), and $S=1/2$ if their orientations are random. It turns out that changing the uptake rate (and hence thickness $h$) from $k=1.6$ to $k=2.8$ changes $S$ by a small amount from $S=0.77$ to $S=0.70$. Here we are more interested in other factors that do not affect $h$.

{\bf Friction.}
One such factor is the nature of friction between cells and the substrate. So far, in all simulations the friction force was proportional to the cell's velocity, irrespective of the direction of motion. To test whether this assumption affected front roughness and the surfing probability, we ran simulations in which friction coefficients were different in the directions parallel and perpendicular to the cell's axis. We replaced Eq. (\ref{eq:drdt}) for the dynamics of the centre of mass with the following equation:
\bq
	\frac{d\vec{r}_i}{dt} = K^{-1} \vec{F}/m,
\eq
where the matrix $K$ accounts for the anisotropy of friction:
\bq
	K=\left[\begin{array}{cc} 
		\zeta_\| n_x^2+\zeta_\bot n_y^2 & (\zeta_\|- \zeta_\bot) n_x n_y  \\
		(\zeta_\|- \zeta_\bot) n_x n_y & \zeta_\bot n_x^2+\zeta_\| n_y^2 
	\end{array}\right].
\eq
We now have two friction coefficients: $\zeta_\bot$ is the coefficient in the direction perpendicular to cell's major axis $\vec{n}$, whereas $\zeta_\|$ is the coefficient in the parallel direction. For convenience, we shall assume that $\zeta_\| = A\zeta, \zeta_\bot = \zeta/A$ where $A$ is the ``asymmetry coefficient'' and $\zeta$ is the isotropic friction coefficient, same as in previous simulations (Table 1).
For isotropic friction, $A=1$, hence $\zeta_\bot = \zeta_\| \equiv \zeta$ and $K=\mathbf{1}\zeta$, and we recover Eq. (\ref{eq:drdt}). If $A>1$, it is easier for the rod to ``roll'' than to slide along the major axis. If $A<1$ it is easier for the rod to slide.

\begin{figure}
\includegraphics{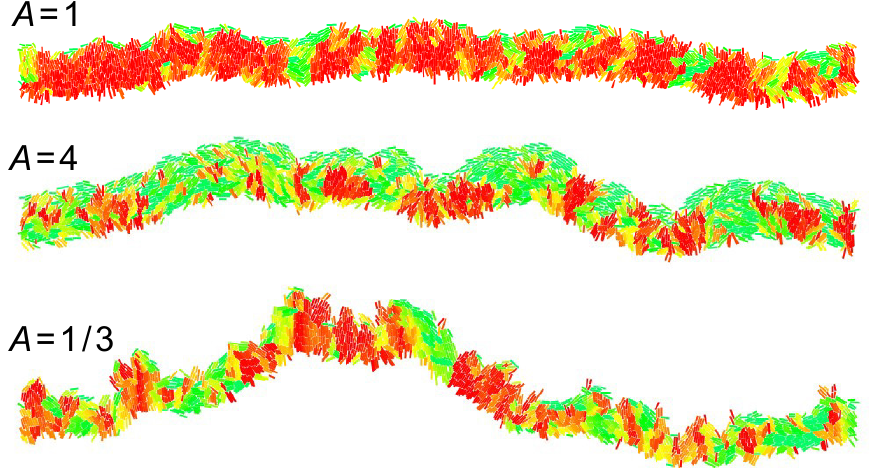}
\caption{\label{anisotropy}Snaphots of a growing colony with different friction anisotropy. The global order parameter $S=0.79$ (isotropic friction $A=1$), $S=0.53$ (rolling rods $A=4$), and $S=0.63$ (sliding rods $A=1/3$).} 
\end{figure}

\begin{figure}
\centering\includegraphics[width=\columnwidth]{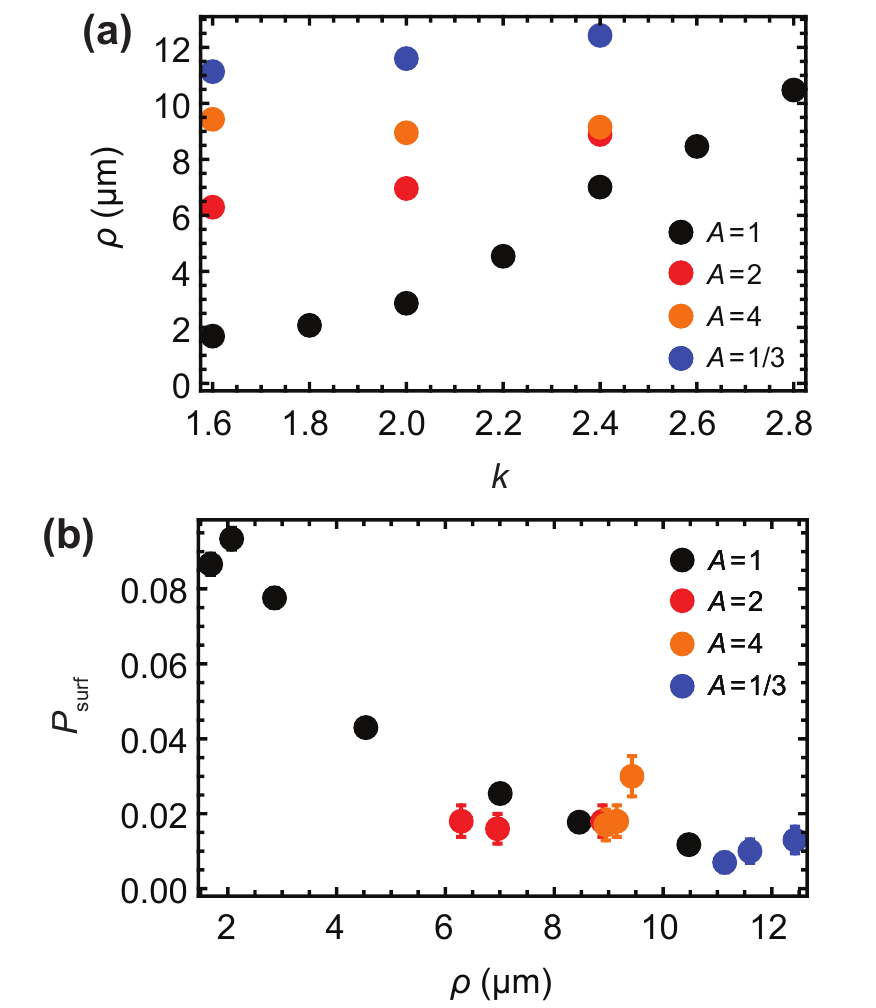}
\caption{\label{aniso} (a) Roughness $\rho$ as the function of $k$, for different levels of friction anisotropy: no anisotropy (black points, $A=1$), ``rolling rods'' $A=2$ (red), $A=4$ (orange), and ``sliding rods'' $A=1/3$ (blue). (b) surfing probability versus $\rho$ for the same parameters as in the left panel.}
\end{figure}

Figure~\ref{anisotropy} shows images of the front for different levels of friction anisotropy. In the anisotropic ``rolling rods'' case ($A>1$), cells are significantly more oriented edge-on to the colony, and the roughness is noticeably larger. In the ``sliding rods'' case ($A<1$) the roughness is even larger but the orientation of cells falls between the isotropic and the ``rolling rods'' case. This is quantified in Fig. \ref{aniso}, left where we plotted $\rho$ as a function of $k$. The same figure, right, shows that, as expected, the surfing probability goes down with increasing roughness.

\section{Comparison with experiments}

\begin{figure*}
 \centering
  \includegraphics[width=0.85\textwidth]
  {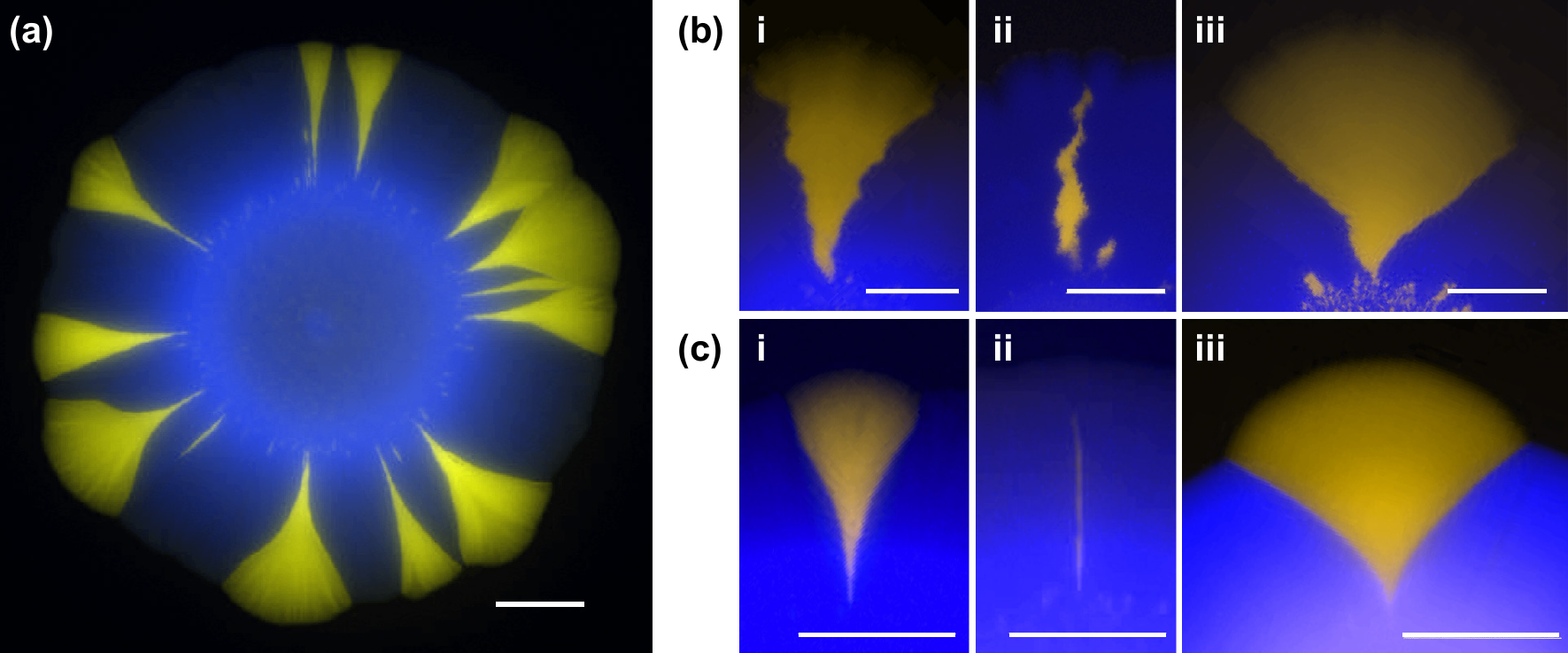}
 \caption{{\bf (a)} An example of a {\it S. cerevisiae} colony with beneficial mutants (yellow) forming sectors. The mutants have a growth rate advantage of $s\approx 10\%$. {\bf (b,c)} Fate of mutant cells - experimental counterpart of Fig.~\ref{sector_shots}. Colonies of {\it E. coli} (b) and {\it S. cerevisiae} (c) were inoculated using a mixture of a majority of wild-type cells (blue, false colour) and a small number of mutant cells (yellow) with $s=8\%$ (left and middle). Some mutant clones formed large sectors (left), while others (middle) lagged behind the front, became engulfed by wild-type cells and eventually ceased to grow ("bubbles"). A large growth advantage ($s\approx 16\%$, right) caused the sector to ``bulge out''. All three phenomena are well reproduced by our simulations (c.f. Fig.~\ref{sector_shots}). In all panels, scale bar = 2mm.
 }\label{fig:expimages}
\end{figure*}

\begin{figure*}
\centering
 \includegraphics
 {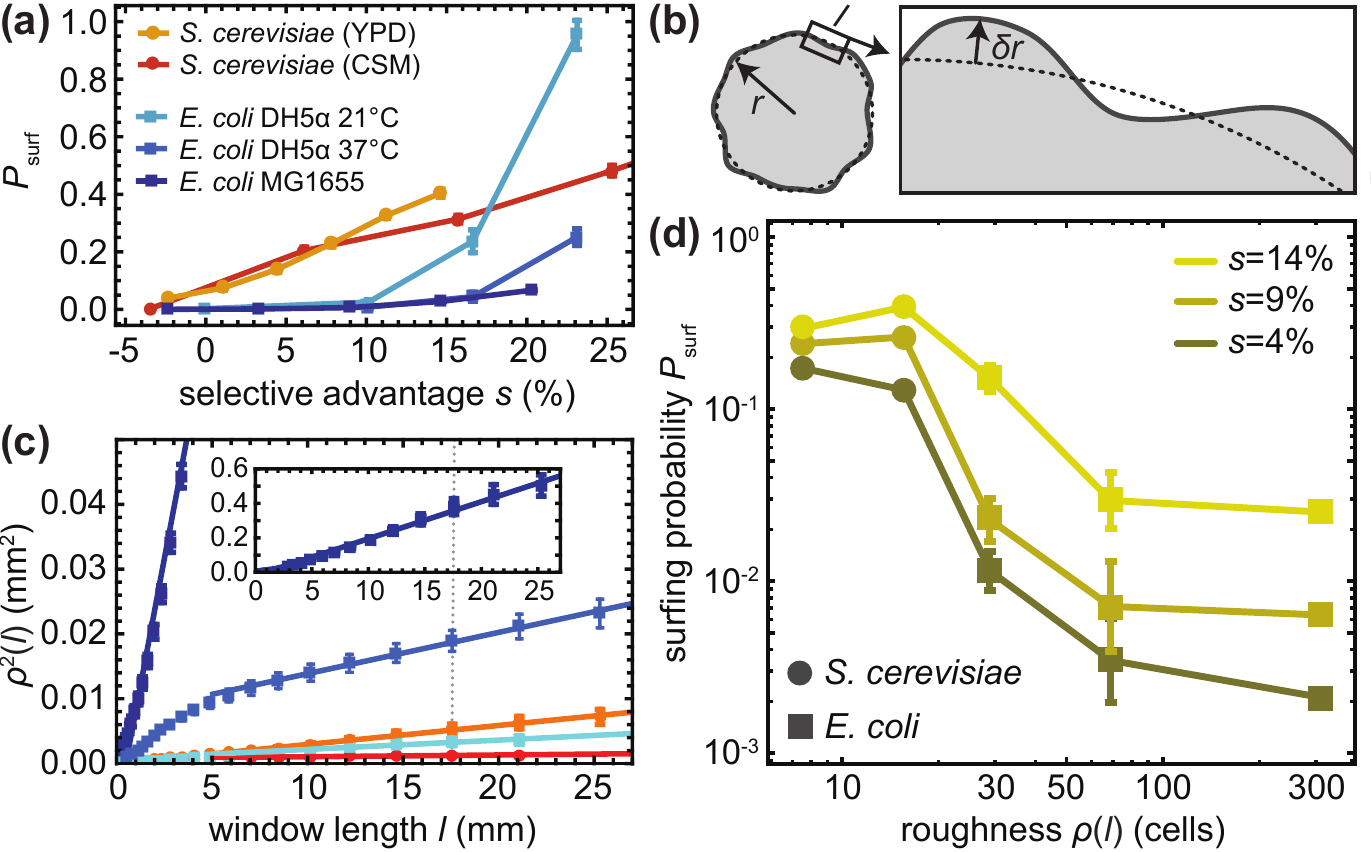}
\caption{Surfing probability versus roughness in experimental colonies. In all panels squares and circles correspond to \textit{E. coli} and \textit{S. cerevisiae}, respectively. \textbf{(a)} Surfing probability $P_{\rm surf}$ for different species and growth conditions as a function of the selective advantage $s$. {\it S. cerevisiae} has a much higher $P_{\rm surf}$ at low $s$, while $P_{\rm surf}$ of \textit{E. coli} strain DH5$\alpha$ at 21C increases faster than linearly for large $s$, surpassing \textit{S. cerevisiae} for $s> 15\%$. \textbf{(b)} Diagram illustraing how roughness $\rho(l)$ was measured (Methods). \textbf{(c)} $\rho^2(l)$ for different conditions (colours as in (a), error bars are standard errors of the mean over at least 10 colonies per condition). Solid lines are linear fits to the data points. The dotted line corresponds to the window length $l=17$mm used to calculate roughness in panel (d). The inset shows $\rho^2(l)$ for \textit{E. coli} MG1655 (dark blue), which has the highest roughness. \textbf{(d)} Surfing probability versus $\rho(l=17{\rm mm})$, for different $s$. To compare \textit{E. coli} and \textit{S. cerevisiae}, we normalized roughness by the cell size (square root of the average area), which we estimated from microscopy images to be 2 and 4.7$\mu$m, respectively.  
}\label{fig:exp1}
\end{figure*}

\begin{figure*}
 \centering\includegraphics
 {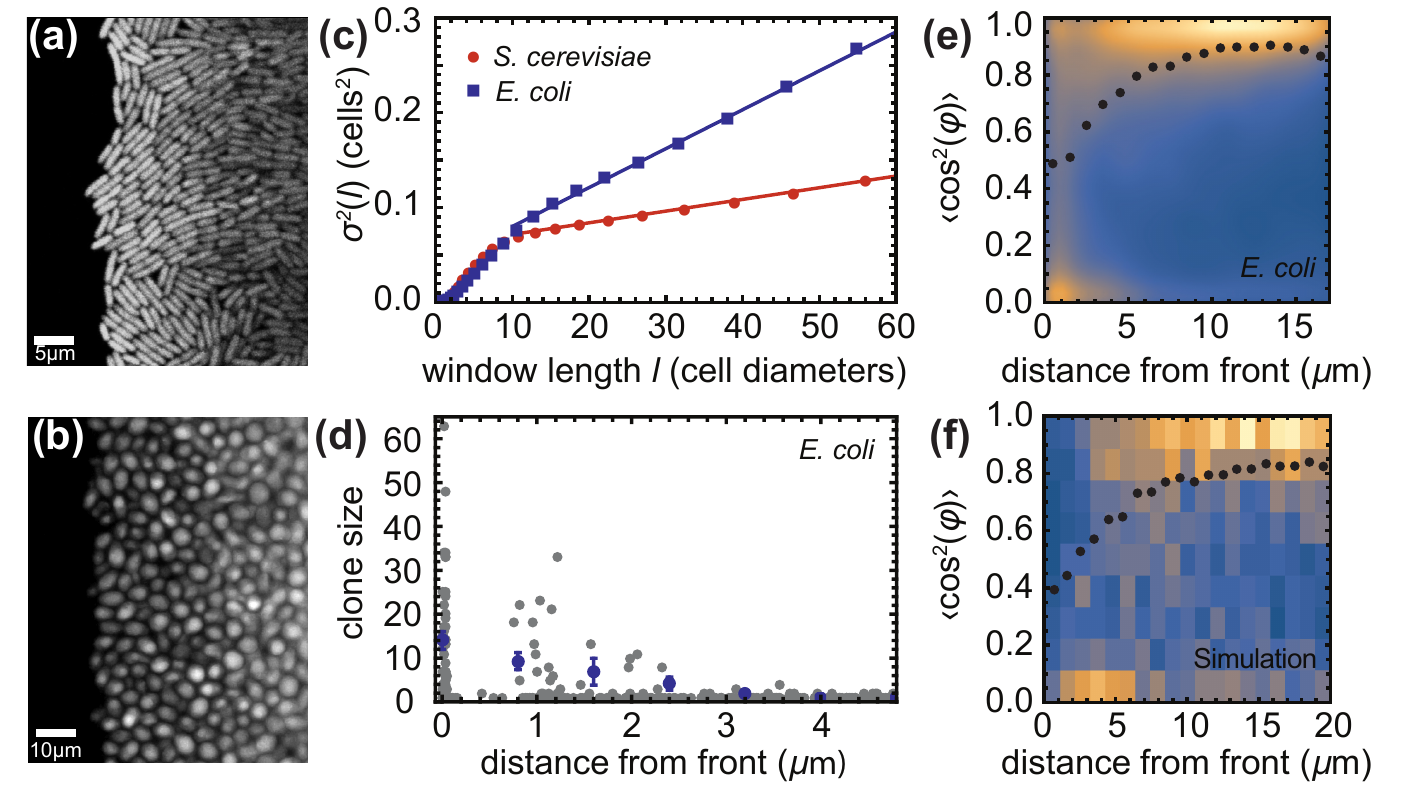}
\caption{Microscopic properties of the growing layer. {\bf (a,b)} Snapshot of an \textit{E. coli} front (panel a, scalebar 5$\mu$m) and a \textit{S. cerevisiae} front (panel b, scalebar 10$\mu$m) front. {\bf (c)} Local roughness $\rho^2(l)$  as a function of the window size $l$. Dashed lines are fits to the data points. {\bf (d)} The number of offspring for all initial cells near the front, for \textit{E. coli}. Only cells within 2-3$\mu$m ($\sim$ one cell) from the edge of the colony have a significant number of offspring. {\bf (e)} Probability density plot of the order parameter $\left<\cos^2\phi\right>$ for \textit{E. coli} as a function of the distance from the edge. Blue = low probability, yellow = high probability. The dotted line is the average order parameter versus the distance from the front. Cells are preferentially aligned with the direction of propagation, except for cells directly at the front, which are parallel to it. {\bf (f)} Density plot of the order parameter for a simulated front with $k=1.4, L=320\mu$m. 
}
\label{fig:exp2}
\end{figure*}

We next checked whether the predicted dependence of the surfing probability on the roughness of the growing layer agree with experiments. We measured surfing probabilities of beneficial mutants with different selective advantages $s=-5\dots 25\%$ in colonies of \textit{E. coli} and \textit{S. cerevisiae} (Methods) grown at different conditions affecting the roughness of the growing layer. A small number of fluorescently labeled mutant cells was mixed with a much larger number of wild-type cells, and a small droplet of the mixture was used to inoculate a colony on a Petri dish. After a few days, colonies with a characteristic sectoring pattern emerged (Fig.~\ref{fig:expimages}). By zooming into the colony edge we confirmed that some mutants ``surfed'' at the front and expanded into large sectors whereas some mutants did not make it and became trapped as bubbles in the bulk of the colony (Fig.~\ref{fig:expimages}, compare with Fig.~\ref{sector_shots}). 

We counted the number of sectors and estimated the surfing probability $P_\text{surf}$ from the formula~\cite{gralka_allele_2016}:
\bq
	P_\mathrm{surf}=\frac{N_\mathrm{sec}}{2\pi r_0P_\mathrm{i}},
    \label{eq:psurfN}
\eq
where $P_\mathrm{i}$ is the initial fraction of mutant cells in the population and $r_0$ the initial radius of the colony (in units of cell diameters). Note this equation makes sense only if surfing is restricted to the first layer of cells; we have shown that this is true in computer simulations and we shall experimentally validate it later in this section. Fig.~\ref{fig:exp1}a shows $P_\mathrm{surf}$ for \textit{E. coli} and \textit{S. cerevisiae}, and for different conditions. In the limit of low selective advantage $s<10\%$ we are interested here, the surfing probability is highest in colonies of roughly-spherical \textit{S. cerevisiae}, which have rather smooth boundaries, and smallest for the rod-shaped bacterium {\it E. coli}, which are characterized by rough front. This agrees with our predictions (Fig. \ref{fig:domains_and_psurf}), however it does not yet show whether this is due to difference in the cell shape or different thickness/roughness of the growing layer.

To study the connection between surfing and surface roughness, we computed the local roughness $\rho(l)$ as a function of window length $l$ (Fig.~\ref{fig:exp1}b, cf. Eq. (\ref{eq:locsigma}) and Methods) for the same colonies for which we previously calculated $P_{\rm surf}$ (Fig.~\ref{fig:exp1}a). In all cases, $\rho^2(l)$ showed a linear dependence on window length $l$ after a transient at small window lengths, i.e., the colony boundary behaved like a standard random walk (Fig.~\ref{fig:exp1}c). 

We then tested the correlation of colony roughness with surfing probability in a similar way to what we did in computer simulations. In Fig.~\ref{fig:exp1}d, we plot the surfing probability $P_\mathrm{surf}$ as a function of colony roughness measured at one specific window length $l=17$mm (dotted line in Fig.~\ref{fig:exp1}c), for different selective advantages $s$. We observe that the surfing probability of {\it E. coli} decreases with increasing roughness (Fig.~\ref{fig:exp1}d) for all $s$, in good qualitative agreement with our simulations. Similar results are obtained for different choices of the window length $l$ for which roughness is calculated. The situation is less clear for \textit{S. cerevisiae}; we hypothesize that this is due to roughness being too small (c.f. Fig. \ref{fig:pfix_rho}) to markedly affect the surfing probability.

We next examined how microscopic properties of the front (cellular orientation) correlated with macroscopic roughness. We analysed microscopic images of the fronts of \textit{E. coli} and \textit{S. cerevisiae} fronts (Methods, data from Ref.~\cite{gralka_allele_2016}), and measured local roughness $\rho(l)$ over sub-mm length scales $l$. Example snapshots in Fig. \ref{fig:exp2}a,b show that roughness of the fronts indeed differ very much for these two microorganisms. Figure \ref{fig:exp2}c confirms that {\it E. coli} has a much higher roughness compared to \textit{S. cerevisiae}, suggesting that macroscopic roughness on the colony scale is a consequence of microscopic front roughness on the single-cell level. 

To study the dynamics of surfing, we tracked \textit{E. coli} cells over 200 minutes and measured their distance from, and orientation relative to the edge of the colony, as well as the number of offspring for all cells in the initial image. Figure ~\ref{fig:exp2}d shows that cells only have an appreciable number of offspring if they are within about one cell diameter of the front. This agrees with our conclusion from simulations and justifies inserting mutants only directly at the front. 

Figure~\ref{fig:exp2}e shows the order parameter $S=\left<\cos^2(\phi-\Phi)\right>$, which measures the orientation of cells and has been defined in Sec. \ref{sec:aniso}, as a function of the distance from the front. Cells near the front tend to align parallel to the front. This changes quickly behind the front, with most cells being perpendicular to the growth direction starting about 5$\mu$m behind the front. Figure~\ref{fig:exp2}f shows the distribution of $S$ obtained from simulations; the agreement with the experimental data from Fig.~\ref{fig:exp2}e is excellent, suggesting that our model indeed captures the dynamics of the growing bacterial front reasonably well.

\section{Conclusions}
In this work we have focused on the role of mechanical interactions in microbial colonies. We first used computer simulations to show that the speed of biological evolution, measured by the probability that a new mutation ``surfs'' at the growing edge of a microbial colony, depends mostly on the thickness and roughness of the growing layer of cells at colony's front. Thicker fronts decrease the per-cell surfing probability because only cells from the very first layer of cells create successful progenies, and the fraction of such cells decreases with increasing front thickness. Rougher fronts also decrease the surfing probability for a similar reason; only cells at the tips of the ``bumps'' are successful and these tips become smaller for rougher fronts. Moreover, roughness and thickness are related; thicker front have lower roughness and vice versa. While the dependence between genetic segregation and the front thickness \cite{nadell_emergence_2010}, and between thickness and roughness \cite{head_linear_2012} has been known previously, in this work we have shown that it is actually the roughness of the growing layer that should be thought of as affecting the surfing probability in the causal sense. We have also linked thickness and roughness to the mechanical properties of cells for the first time. Moreover, we have discovered that the orientation of cells has also a significant effect, irrespective of front roughness, on the surfing probability.

All these quantities (roughness, thickness, cellular alignment) are controlled in a very non-trivial way by the properties of cells and their environment: cell-surface friction (and anisotropy of thereof), elasticity of cells, their growth/nutrient uptake rate, and their shape. While some of these parameters are very difficult to vary experimentally, we managed to show in simple experiments that the growth rate and the shape of cells affect the surfing probability in the way predicted by our simulations.

Microbial evolution is a research area that is important both from fundamental and practical viewpoints. In particular, our research shows that mechanical forces such as adhesion, friction, etc., can play a significant role in biological evolution of microorganisms. To our knowledge, this article is the first that not only puts forward this idea but also provides concrete arguments in its support.

From a more practical point of view, our results are relevant to the evolution of antimicrobial resistance. It has been demonstrated that even a small bacterial population can develop {\it de novo} resistance to some antimicrobial drugs in less than a day \cite{zhang_acceleration_2011}. This rapid evolution makes the most popular drugs - antibiotics - increasingly ineffective \cite{baquero_evolution_1997}. Since the rate of discovery of new antibiotics has steadily declined over years \cite{spellberg_epidemic_2008}, the evolution of drug-resistant bacteria has been highlighted as one of the major challenges we will face in the coming decades. By demonstrating the role of mechanical interactions on biological evolution in microbial aggregates, our research opens up a new antimicrobial paradigm in which the physical properties of microbes could be targeted alongside standard antimicrobial therapy to reduce the probability of evolving resistance to drugs. 

\section*{Acknowledgments}
Research reported in this publication was supported by the Royal Society of Edinburgh (B.W.), National Institute of General Medical Sciences of the National Institutes of Health under Award Number R01GM115851 (O.H.), by a National Science Foundation Career Award (O.H.) and by a Simons Investigator award from the Simons Foundation (O.H.). The content is solely the responsibility of the authors and does not necessarily represent the official views of the National Institutes of Health.

\bibliographystyle{verbose}

\bibliography{library}

\begin{thebibliography}{10}

\bibitem{schlegel_general_1993}
H.~G. Schlegel, C.~Zaborosch, and M.~Kogut.
\newblock \emph{General {Microbiology}}.
\newblock Cambridge University Press, 1993.
\newblock ISBN 978-0-521-43980-0.

\bibitem{pieper_engineering_2000}
D.~H. Pieper and W.~Reineke.
\newblock Engineering bacteria for bioremediation.
\newblock \emph{Current Opinion in Biotechnology}, \textbf{11}(3):262--270,
  2000.

\bibitem{sabra_biosystems_2010}
W.~Sabra, D.~Dietz, D.~Tjahjasari, and A.-P. Zeng.
\newblock Biosystems analysis and engineering of microbial consortia for
  industrial biotechnology.
\newblock \emph{Engineering in Life Sciences}, \textbf{10}(5):407--421, 2010.

\bibitem{chattopadhyay_high_2009}
S.~Chattopadhyay, S.~J. Weissman, V.~N. Minin, T.~A. Russo, D.~E. Dykhuizen,
  and E.~V. Sokurenko.
\newblock High frequency of hotspot mutations in core genes of {Escherichia}
  coli due to short-term positive selection.
\newblock \emph{Proceedings of the National Academy of Sciences},
  \textbf{106}(30):12412--12417, 2009.

\bibitem{koch_evolution_2014}
G.~Koch, A.~Yepes, K.~F\"{o}rstner, C.~Wermser, S.~Stengel, J.~Modamio,
  K.~Ohlsen, K.~Foster, and D.~Lopez.
\newblock Evolution of {Resistance} to a {Last}-{Resort} {Antibiotic} in
  {Staphylococcus} aureus via {Bacterial} {Competition}.
\newblock \emph{Cell}, \textbf{158}(5):1060--1071, 2014.

\bibitem{elena_evolution_2003}
S.~F. Elena and R.~E. Lenski.
\newblock Evolution experiments with microorganisms: the dynamics and genetic
  bases of adaptation.
\newblock \emph{Nature Reviews Genetics}, \textbf{4}(6):457--469, 2003.

\bibitem{perron_rate_2008}
G.~G. Perron, A.~Gonzalez, and A.~Buckling.
\newblock The rate of environmental change drives adaptation to an antibiotic
  sink.
\newblock \emph{Journal of Evolutionary Biology}, \textbf{21}(6):1724--1731,
  2008.

\bibitem{carpentier_biofilms_1993}
B.~Carpentier and O.~Cerf.
\newblock Biofilms and their consequences, with particular reference to hygiene
  in the food industry.
\newblock \emph{Journal of Applied Microbiology}, \textbf{75}(6):499--511,
  1993.

\bibitem{costerton_bacterial_1999}
J.~W. Costerton, P.~S. Stewart, and E.~P. Greenberg.
\newblock Bacterial {Biofilms}: {A} {Common} {Cause} of {Persistent}
  {Infections}.
\newblock \emph{Science}, \textbf{284}(5418):1318--1322, 1999.

\bibitem{berry_microbial_2006}
D.~Berry, C.~Xi, and L.~Raskin.
\newblock Microbial ecology of drinking water distribution systems.
\newblock \emph{Current Opinion in Biotechnology}, \textbf{17}(3):297--302,
  2006.

\bibitem{gibson_pathophysiology_2003}
R.~L. Gibson, J.~L. Burns, and B.~W. Ramsey.
\newblock Pathophysiology and {Management} of {Pulmonary} {Infections} in
  {Cystic} {Fibrosis}.
\newblock \emph{American Journal of Respiratory and Critical Care Medicine},
  \textbf{168}(8):918--951, 2003.

\bibitem{stewart_antibiotic_2001}
P.~S. Stewart and J.~William~Costerton.
\newblock Antibiotic resistance of bacteria in biofilms.
\newblock \emph{The Lancet}, \textbf{358}(9276):135--138, 2001.

\bibitem{drenkard_antimicrobial_2003}
E.~Drenkard.
\newblock Antimicrobial resistance of {Pseudomonas} aeruginosa biofilms.
\newblock \emph{Microbes and Infection}, \textbf{5}(13):1213--1219, 2003.

\bibitem{breidenstein_pseudomonas_2011}
E.~B.~M. Breidenstein, C.~de~la Fuente-Nunez, and R.~E.~W. Hancock.
\newblock Pseudomonas aeruginosa: all roads lead to resistance.
\newblock \emph{Trends in Microbiology}, \textbf{19}(8):419--426, 2011.

\bibitem{singh_quorum_sensing_2000}
P.~K. Singh, A.~L. Schaefer, M.~R. Parsek, T.~O. Moninger, M.~J. Welsh, and
  E.~P. Greenberg.
\newblock Quorum-sensing signals indicate that cystic fibrosis lungs are
  infected with bacterial biofilms.
\newblock \emph{Nature}, \textbf{407}(6805):762--764, 2000.

\bibitem{boyer_buckling_2011}
D.~Boyer, W.~Mather, O.~Mondragon-Palomino, S.~Orozco-Fuentes, T.~Danino,
  J.~Hasty, and L.~S. Tsimring.
\newblock Buckling instability in ordered bacterial colonies.
\newblock \emph{Physical Biology}, \textbf{8}(2):026008, 2011.

\bibitem{Farrell2013}
F.~D.~C. Farrell, O.~Hallatschek, D.~Marenduzzo, and B.~Waclaw.
\newblock {Mechanically Driven Growth of Quasi-Two-Dimensional Microbial
  Colonies}.
\newblock \emph{Phys. Rev. Lett.}, \textbf{111}(16):168101, 2013.

\bibitem{giverso_emerging_2015}
C.~Giverso, M.~Verani, and P.~Ciarletta.
\newblock Emerging morphologies in round bacterial colonies: comparing
  volumetric versus chemotactic expansion.
\newblock \emph{Biomechanics and Modeling in Mechanobiology}, 2015.

\bibitem{ghosh_mechanically_driven_2015}
P.~Ghosh, J.~Mondal, E.~Ben-Jacob, and H.~Levine.
\newblock Mechanically-driven phase separation in a growing bacterial colony.
\newblock \emph{Proceedings of the National Academy of Sciences}, page
  201504948, 2015.

\bibitem{Volfson2008}
D.~Volfson, S.~Cookson, J.~Hasty, and L.~S. Tsimring.
\newblock {Biomechanical ordering of dense cell populations.}
\newblock \emph{Proc. Natl. Acad. Sci. U. S. A.}, \textbf{105}(40):15346--51,
  2008.

\bibitem{su_bacterial_2012}
P.-T. Su, C.-T. Liao, J.-R. Roan, S.-H. Wang, A.~Chiou, and W.-J. Syu.
\newblock Bacterial {Colony} from {Two}-{Dimensional} {Division} to
  {Three}-{Dimensional} {Development}.
\newblock \emph{PLoS ONE}, \textbf{7}(11):e48098, 2012.

\bibitem{asally_cover_2012}
M.~Asally, M.~Kittisopikul, P.~Rue, Y.~Du, Z.~Hu, T.~Cagatay, A.~B. Robinson,
  H.~Lu, J.~Garcia-Ojalvo, and G.~M. Suel.
\newblock From the {Cover}: {Localized} cell death focuses mechanical forces
  during 3d patterning in a biofilm.
\newblock \emph{Proceedings of the National Academy of Sciences},
  \textbf{109}(46):18891--18896, 2012.

\bibitem{grant_role_2014}
M.~A.~A. Grant, B.~Wacaw, R.~J. Allen, and P.~Cicuta.
\newblock The role of mechanical forces in the planar-to-bulk transition in
  growing {Escherichia} coli microcolonies.
\newblock \emph{Journal of The Royal Society Interface},
  \textbf{11}(97):20140400--20140400, 2014.

\bibitem{oldewurtel_differential_2015}
E.~R. Oldewurtel, N.~Kouzel, L.~Dewenter, K.~Henseler, B.~Maier, and R.~Kolter.
\newblock Differential interaction forces govern bacterial sorting in early
  biofilms.
\newblock \emph{eLife}, \textbf{4}:e10811, 2015.

\bibitem{klopfstein_fate_2006}
S.~Klopfstein, M.~Currat, and L.~Excoffier.
\newblock The {Fate} of {Mutations} {Surfing} on the {Wave} of a {Range}
  {Expansion}.
\newblock \emph{Molecular Biology and Evolution}, \textbf{23}(3):482--490,
  2006.

\bibitem{fusco2016excess}
D.~Fusco, M.~Gralka, A.~Anderson, J.~Kayser, and O.~Hallatschek.
\newblock Excess of mutational jackpot events in growing populations due to
  gene surfing.
\newblock \emph{bioRxiv}, page 053405, 2016.

\bibitem{Hallatschek2007}
O.~Hallatschek, P.~Hersen, S.~Ramanathan, and D.~R. Nelson.
\newblock {Genetic drift at expanding frontiers promotes gene segregation.}
\newblock \emph{Proc. Natl. Acad. Sci. U. S. A.}, \textbf{104}(50):19926--30,
  2007.

\bibitem{excoffier_genetic_2009}
L.~Excoffier, M.~Foll, and R.~J. Petit.
\newblock Genetic consequences of range expansions.
\newblock \emph{Annual Review of Ecology, Evolution, and Systematics},
  \textbf{40}:481--501, 2009.

\bibitem{Hallatschek2010}
O.~Hallatschek and D.~R. Nelson.
\newblock {Life at the front of an expanding population.}
\newblock \emph{Evolution}, \textbf{64}(1):193--206, 2010.

\bibitem{behrman_species_2011}
K.~D. Behrman and M.~Kirkpatrick.
\newblock Species range expansion by beneficial mutations.
\newblock \emph{Journal of Evolutionary Biology}, \textbf{24}(3):665--675,
  2011.

\bibitem{ali_reproduction_time_2012}
A.~Ali, E.~Somfai, and S.~Grosskinsky.
\newblock Reproduction-time statistics and segregation patterns in growing
  populations.
\newblock \emph{Physical Review E}, \textbf{85}(2):021923, 2012.

\bibitem{Korolev2012}
K.~S. Korolev, M.~J.~I. M\"{u}ller, N.~Karahan, A.~W. Murray, O.~Hallatschek,
  and D.~R. Nelson.
\newblock {Selective sweeps in growing microbial colonies.}
\newblock \emph{Phys. Biol.}, \textbf{9}(2):026008, 2012.

\bibitem{Lehe2012}
R.~Lehe, O.~Hallatschek, and L.~Peliti.
\newblock {The rate of beneficial mutations surfing on the wave of a range
  expansion.}
\newblock \emph{PLoS Comput. Biol.}, \textbf{8}(3):e1002447, 2012.

\bibitem{gralka_allele_2016}
M.~Gralka, F.~Stiewe, F.~Farrell, W.~M\"{o}bius, B.~Waclaw, and O.~Hallatschek.
\newblock Allele surfing promotes microbial adaptation from standing variation.
\newblock \emph{Ecology Letters}, pages n/a--n/a, 2016.

\bibitem{murray}
J.~D. Murray.
\newblock \emph{{Mathematical Biology, Vol. 2}}.
\newblock Springer-Verlag, Berlin, 2003.

\bibitem{kreft_biofilms_2004}
J.-U. Kreft.
\newblock Biofilms promote altruism.
\newblock \emph{Microbiology}, \textbf{150}(8):2751--2760, 2004.

\bibitem{Xavier2007}
J.~B. Xavier and K.~R. Foster.
\newblock {Cooperation and conflict in microbial biofilms}.
\newblock \emph{Proc. Natl. Acad. Sci.}, \textbf{104}(3):876--881, 2007.

\bibitem{Xavier2009}
J.~a.~B. Xavier, E.~Martinez-Garcia, and K.~R. Foster.
\newblock {Social evolution of spatial patterns in bacterial biofilms: when
  conflict drives disorder.}
\newblock \emph{Am. Nat.}, \textbf{174}(1):1--12, 2009.

\bibitem{hoffman_synchrony_1965}
H.~Hoffman and M.~E. Frank.
\newblock Synchrony of division in clonal microcolonies of {Escherichia} coli.
\newblock \emph{Journal of bacteriology}, \textbf{89}(2):513--517, 1965.

\bibitem{kennard_individuality_2016}
A.~S. Kennard, M.~Osella, A.~Javer, J.~Grilli, P.~Nghe, S.~J. Tans, P.~Cicuta,
  and M.~Cosentino~Lagomarsino.
\newblock Individuality and universality in the growth-division laws of single
  {\textbackslash}textit\{{E}. coli\} cells.
\newblock \emph{Physical Review E}, \textbf{93}(1):012408, 2016.

\bibitem{iyer_biswas_scaling_2014}
S.~Iyer-Biswas, C.~S. Wright, J.~T. Henry, K.~Lo, S.~Burov, Y.~Lin, G.~E.
  Crooks, S.~Crosson, A.~R. Dinner, and N.~F. Scherer.
\newblock Scaling laws governing stochastic growth and division of single
  bacterial cells.
\newblock \emph{Proceedings of the National Academy of Sciences},
  \textbf{111}(45):15912--15917, 2014.

\bibitem{freese_genetic_2014}
P.~Freese, K.~Korolev, J.~Jimenez, and I.~Chen.
\newblock Genetic {Drift} {Suppresses} {Bacterial} {Conjugation} in {Spatially}
  {Structured} {Populations}.
\newblock \emph{Biophysical Journal}, \textbf{106}(4):944--954, 2014.

\bibitem{Fujikawa1989}
H.~Fujikawa and M.~Matsushita.
\newblock {Fractal Growth of Bacillus subtilis on Agar Plate}.
\newblock \emph{J. Phys. Soc. Jpn}, 1989.

\bibitem{Kawasaki1997}
K.~Kawasaki, A.~Mochizuki, M.~Matsushita, T.~Umeda, and N.~Shigesada.
\newblock {Modeling spatio-temporal patterns generated by Bacillus subtilis.}
\newblock \emph{J. Theor. Biol.}, \textbf{188}(2):177--85, 1997.

\bibitem{M.A.Nowak2006}
{M. A. Nowak}.
\newblock \emph{{Evolutionary Dynamics}}.
\newblock Belknap/Harvard, Camridge, Massachusetts, 2006.

\bibitem{nadell_emergence_2010}
C.~D. Nadell, K.~R. Foster, and J.~B. Xavier.
\newblock Emergence of {Spatial} {Structure} in {Cell} {Groups} and the
  {Evolution} of {Cooperation}.
\newblock \emph{PLoS Computational Biology}, \textbf{6}(3):e1000716, 2010.

\bibitem{head_linear_2012}
D.~A. Head.
\newblock Linear surface roughness growth and flow smoothening in a
  three-dimensional biofilm model.
\newblock \emph{arXiv preprint arXiv:1210.8103}, 2012.

\bibitem{zhang_acceleration_2011}
Q.~Zhang, G.~Lambert, D.~Liao, H.~Kim, K.~Robin, C.-k. Tung, N.~Pourmand, and
  R.~H. Austin.
\newblock Acceleration of {Emergence} of {Bacterial} {Antibiotic} {Resistance}
  in {Connected} {Microenvironments}.
\newblock \emph{Science}, \textbf{333}(6050):1764--1767, 2011.

\bibitem{baquero_evolution_1997}
F.~Baquero and J.~Blazquez.
\newblock Evolution of antibiotic resistance.
\newblock \emph{Trends in Ecology \& Evolution}, \textbf{12}(12):482--487,
  1997.

\bibitem{spellberg_epidemic_2008}
B.~Spellberg, R.~Guidos, D.~Gilbert, J.~Bradley, H.~W. Boucher, W.~M. Scheld,
  J.~G. Bartlett, J.~Edwards, and {the Infectious Diseases Society of America}.
\newblock The {Epidemic} of {Antibiotic}-{Resistant} {Infections}: {A} {Call}
  to {Action} for the {Medical} {Community} from the {Infectious} {Diseases}
  {Society} of {America}.
\newblock \emph{Clinical Infectious Diseases}, \textbf{46}(2):155--164, 2008.

\end{thebibliography}

\end{document}